\documentclass{iopart}

\usepackage{times}
\usepackage{graphicx}
\usepackage{iopams}

\renewcommand{\vec}[1]{\boldsymbol{#1}}
\renewcommand{\d}{\rmd}
\renewcommand{\i}{\rmi}
\renewcommand{\mat}[1]{\boldsymbol{#1}}
\newcommand{\mutt}{\mat{\mu}^{\rm tt}}
\newcommand{\murr}{\mat{\mu}^{\rm rr}}
\newcommand{\mutr}{\mat{\mu}^{\rm tr}}
\newcommand{\murt}{\mat{\mu}^{\rm rt}}
\newcommand{\mut}{\mu^{\rm t}}

\begin{document}

%%% TITLE %%%%%%%%%%%%%%%%%%%%%%%%%%%%%%%%%%%%%%%%%%%%%%%%%%%%%%%%%%%%%%%%%%%%%

\vspace*{-2.6cm}

\title{Circling particles and drafting in optical vortices}
\author{Michael Reichert and Holger Stark}
\address{Fachbereich Physik, Universit\"at Konstanz,
  D-78457 Konstanz, Germany}
\ead{michael.reichert@uni-konstanz.de}

\begin{abstract}
Particles suspended in a viscous fluid circle in optical vortices
generated by holographic optical-tweezer techniques [Curtis J E and
Grier D G 2003 {\it Phys. Rev. Lett.} {\bf 90} 133901]. We model this
system and show that hydrodynamic interactions between the circling
particles determine their collective motion. We perform a
linear-stability analysis to investigate the stability of regular
particle clusters and illustrate the limit cycle to which the unstable
modes converge. We clarify that drafting of particle doublets is
essential for the understanding of the limit cycle.
\end{abstract}

\vspace{-5mm}


%%% INTRODUCTION %%%%%%%%%%%%%%%%%%%%%%%%%%%%%%%%%%%%%%%%%%%%%%%%%%%%%%%%%%%%%%

\section{Introduction}
\label{sec:introduction}

Hydrodynamic interactions occur between particles or bodies whenever
they move relative to each other in a viscous fluid. Due to their
long-range nature, they are important for the dynamic properties
of colloidal suspensions exemplified by the self- and collective 
diffusion \cite{pus91,dho96,naeg96,ban00}, sedimentation
\cite{lad93,bre99,fel03}, or the aggregation of particles
\cite{tan00}. Through the correlated motion of a pair of colloids
trapped in optical tweezers, one can directly measure the effect of
hydrodynamic interactions \cite{mei99,bar01,hen01,rei04}.
Furthermore, hydrodynamic interactions give rise to interesting
collective behavior, e.g., periodic or almost periodic motions in time
\cite{caf88,sno97} or even transient chaotic dynamics in 
the sedimentation of particles \cite{jan97}. And they lead to 
pattern formation of rotating motors \cite{grzyb00} 
with a possible 2D melting transition of biological motors like 
ATP-synthase embedded in a membrane \cite{len03}. 
Hydrodynamic interactions are treated in the low-Reynolds-number
regime which is also relevant for biology. It determines the problem 
of how microorganisms move forward \cite{pur77}. Certain 
bacteria accomplish this, e.g., by cranking helical flexible rods
\cite{pow03}.
In addition, recent experiments show that laminar flows initiate the 
asymmetries between the right- and left-hand side of the body at an
early stage of the embryonic development \cite{ster02,ess02,non02}.

The work presented here is motivated by experiments of
Curtis and Grier \cite{cur03}. They created toroidal optical
traps known as optical vortices with the help of holographic
techniques. In the bright circumference of an optical vortex, they
could trap particles which circulated around the ring due to
scattering forces which result from the orbital angular momentum
of light. In this article, we model the system of circling
particles. We demonstrate that hydrodynamic interactions
determine their interesting collective motion which we analyze by 
methods used in the study of non-linear dynamics. In particular, we
identify a limit cycle which is governed by drafting of particle
doublets.

The outline of the article is as follows. In \sref{sec:model}, 
we first define our model system of particles moving in a ring-like
trap and summarize the basic equations for the description of the
dynamics. The stability of regular $N$-particle clusters in the ring
is investigated in \sref{sec:stability} by means of a linear 
stability analysis. \Sref{sec:nonlinear_dynamics} then describes 
the nonlinear dynamics of perturbed clusters. We introduce the 
periodic limit cycle of the three-particle system and perform a 
harmonic analysis. The influence of the trap strength on the 
dynamics is discussed. Furthermore, we briefly address the dynamics 
in fairly weak traps. Finally, in \sref{sec:velocities}, we present
the particle velocities as a function of ring radius and particle
number.

%%% MODEL %%%%%%%%%%%%%%%%%%%%%%%%%%%%%%%%%%%%%%%%%%%%%%%%%%%%%%%%%%%%%%%%%%%%%

\section{Modelling particle dynamics in a toroidal optical trap}
\label{sec:model}

\begin{figure}
\begin{indented}
\item
\includegraphics[width=0.63\textwidth]{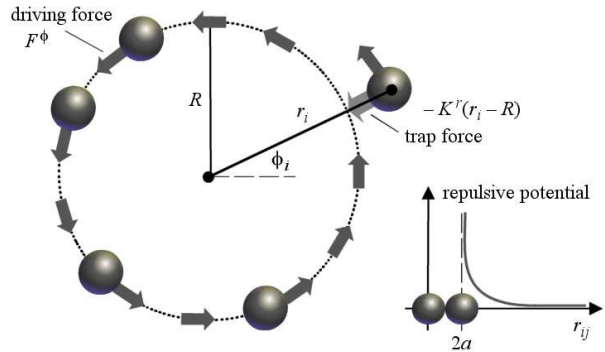}
\end{indented}
\caption{Definition of our model system of an optical vortex. The
particles are driven by a constant tangential force and trapped in the ring
by a harmonic radial force. In order to avoid overlaps, the particles
repel each other at very short distances.}
\label{fig:ring-trap}
\end{figure}

We consider non-Brownian, equal-sized spherical particles suspended in
a viscous fluid in the regime of low Reynolds numbers whose mutual
interactions are solely of hydrodynamic origin.
We mimick the basic features of particles captured in an optical
vortex by applying a constant tangential force $F^{\phi}$ to each
particle and by keeping the particles on a ring of radius $R$ by means
of a harmonic radial trap with force constant $K^{r}$
(\fref{fig:ring-trap}).
The particle motion is effectively two-dimensional in the plane of the
ring ($z=0$); thus, the particle positions are best described by polar
coordinates, $r_{i}$ and $\phi_{i}$. With the radial and tangential
unit vectors at the position of particle $i$,
$\vec{e}_{i}^{r}=(\cos\phi_{i},\sin\phi_{i},0)$ and
$\vec{e}_{i}^{\phi}=(-\sin\phi_{i},\cos\phi_{i},0)$, the total
external force acting on particle $i$ then reads
\begin{equation}
\vec{F}_{i} = F^{\phi}\vec{e}_{i}^{\phi}
- K^{r}(r_{i}-R)\vec{e}_{i}^{r} \, . 
\label{eq:vortex-force}
\end{equation}

Note that the assumption of a constant tangential driving force
$F^{\phi}$ is a rough approximation since the intensity profile of an
optical vortex is modulated along the ring as described by the
topological charge $\ell$ of the vortex\ \cite{cur03}. However,
if $\ell$ is large so that the period of the modulations is
small compared to the particle size, then our approximation
seems to be reasonable.

In order to prevent that particles overlap in the simulations, we add a
repulsive interaction which becomes relevant only when two particles
are very close to each other. We use a hard-sphere-type interaction
potential where a ``hard core'' with diameter $2a$ is surrounded by a
``soft'' repulsive potential (\fref{fig:ring-trap}) \cite{bry02}:
\begin{equation}
V^{\rm rep}(r_{ij}) = W\left[\left(\frac{r_{ij}}{2a}\right)^{m}
-1\right]^{-1} \, .
\label{eq:repulsion}
\end{equation}
Here, $r_{ij}$ is the center-to-center distance between particles $i$
and $j$, and $a$ is the particle radius. This choice is primarily due
to numerical reasons, but can nevertheless be justified from the
physical point of view since even hard-sphere-like colloids show a
``soft'' repulsive potential at very short distances \cite{bry02}.
We choose the commonly used Lennard-Jones exponent $m=12$ so that the
repulsion only acts at very short distances, and we tune the prefactor
$W$ such that the minimal inter-particle gap $r_{ij}-2a$ in the
simulations is of the order of $10^{-4}a$.

In the regime of low Reynolds numbers, the flow of an incompressible
fluid with viscosity $\eta$ obeys the Stokes or creeping flow
equations \cite{bre63,bre64,dho96}
\begin{equation}
\eta\nabla^{2}\vec{u}-\nabla p = \vec{0} \, , \quad
\nabla\cdot\vec{u} = 0 \, ,
\label{eq:stokes}
\end{equation}
where $\vec{u}$ is the flow field and $p$ the hydrodynamic pressure.
Imposing stick boundary conditions on the surfaces of all particles
suspended in the fluid, the motions of the particles are mutually
coupled via the flow field. Due to the linearity of
\eref{eq:stokes}, the translational and rotational velocities of the
particles, $\vec{v}_{i}$ and $\vec{\omega}_{i}$ ($i=1,\ldots,N$),
depend linearly on all external forces and torques acting on the particles,
$\vec{F}_{j}$ and $\vec{T}_{j}$ \cite{bre63,bre64,dho96}:
\begin{equation}
\vec{v}_{i} = \sum\limits_{j=1}^{N}\left(\mutt_{ij}\vec{F}_{j}
+\mutr_{ij}\vec{T}_{j}\right) \, ,
\quad
\vec{\omega}_{i} = \sum\limits_{j=1}^{N}\left(\murt_{ij}\vec{F}_{j}
+\murr_{ij}\vec{T}_{j}\right) \, .
\label{eq:(v,omega)=((mu))*(F,T)}
\end{equation}
The central quantities are the $3\times 3$ mobility
tensors $\mutt_{ij}$, $\murr_{ij}$, $\mutr_{ij}$, and $\murt_{ij}$. They
depend on the current spatial configuration of all particles, i.e.,
the set of position vectors $\{\vec{r}_{1},\ldots,\vec{r}_{N}\}$ in
the case of spherical shape.

As there are no external torques in our model, i.e.,
$\vec{T}_{j}\equiv\vec{0}$, and as we are not interested in the
rotational motion of the particles, the
remaining equation of motion for our problem is
\begin{equation}
\dot{\vec{r}}_{i} \equiv \vec{v}_{i}
= \sum\limits_{j=1}^{N}\mutt_{ij}\vec{F}_{j} \, .
\label{eq:r'=mu*F}
\end{equation}
Since the mobility tensors $\mutt_{ij}$ are nonlinear functions of all
particle positions $\{\vec{r}_{k}\}$, \eref{eq:r'=mu*F} describes the
coupled nonlinear dynamics of $N$ particles.

The first-order approximation (with respect to inverse particle
distances $r_{ij}$) for the mobilities of particles moving in an
unbounded and otherwise quiescent fluid is the well-known Oseen tensor
\cite{dho96}
\begin{equation}
\mutt_{ij}=\frac{3\mut a}{4r_{ij}}\,
\bigg(\mat{1}+\frac{\vec{r}_{ij}\otimes\vec{r}_{ij}}{r_{ij}^{2}}\bigg)
\, ,
\label{eq:oseen}
\end{equation}
where $\mut=(6\pi\eta a)^{-1}$ is the Stokes mobility for a translating
sphere; the self-mobilities are $\mutt_{ii}=\mut\mat{1}$. All the
other mobility tensors in \eref{eq:(v,omega)=((mu))*(F,T)} vanish in
this approximation. Note that the Oseen tensor is the Green function
of the Stokes equations \eref{eq:stokes}, i.e., it considers pointlike
particles and hence does not include rotational couplings.

Higher order approximations of all mobility tensors can be calculated,
e.g., via the multipole expansion method \cite{cich94}. It has been
implemented in the numerical library {\sc hydrolib} \cite{hin95},
which we use in our simulations. For the numerical integration of the
highly nonlinear equation of motion \eref{eq:r'=mu*F}, we apply a
fourth-order Runge-Kutta scheme. The time step is chosen such that the
corresponding change in the angular position is of the order of
$1^{\circ}$.

%%% STABILITY %%%%%%%%%%%%%%%%%%%%%%%%%%%%%%%%%%%%%%%%%%%%%%%%%%%%%%%%%%%%%%%%%

\section{Stability analysis of regular clusters}
\label{sec:stability}

\begin{figure}
\begin{indented}
\item
\includegraphics[width=0.63\textwidth]{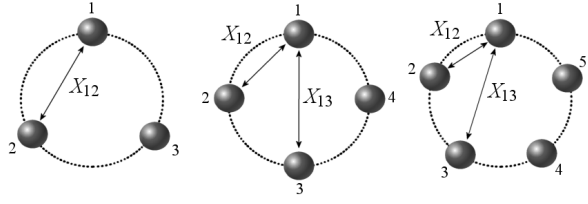}
\end{indented}
\vspace{-3mm}
\caption{Regular $N$-particle clusters. }
\label{fig:clusters}
\end{figure}

In this section, we study the stability of regular $N$-particle
clusters. By ``regular'' we mean configurations with $N$-fold
symmetry, like the ones shown in \fref{fig:clusters}.
Due to this symmetry, all particle positions are equivalent,
i.e., the angular velocity $\dot{\phi}_{i}$ has to be the same for all
particles. The radial positions of the particles do not change either,
which can also be justified by a simple symmetry
argument. Let us start with a regular configuration where $r_{i}=R$ for
all particles, so there is not radial force component in
\eref{eq:vortex-force}. Due to linearity, $\dot{r}_{i}\propto
F^{\phi}$, which means that changing the direction of $F^{\phi}$
inverts the direction of $\dot{r}_{i}$. However, since there
is no difference if the cluster rotates clockwise or counterclockwise,
the radial velocity must be the same for both directions, i.e.,
$\dot{r}_{i}\equiv 0$ and $r_{i}\equiv R$. Therefore, a regular
cluster remains unchanged and rotates with constant frequency
$\Omega_{N}$.

In regular clusters, the particles are well separated for sufficiently
large radius $R$, and we may approximate the hydrodynamic
interactions by the Oseen tensor \eref{eq:oseen}. The particle
positions of the regular clusters are given by $r_{i}=R$ and
$\phi_{i}=\Phi_{i}(t)=(2\pi/N)i+\Omega_{N}t$. With
$\vec{F}_{i}=F^{\phi}\vec{e}_{i}^{\phi}$, we derive from
\eref{eq:r'=mu*F} the rotational frequency of an $N$-particle
cluster,
\begin{equation}
\Omega_{N}=\Omega_{1}\bigg[1+\frac34\sum\limits_{j\neq i}
\frac{1}{X_{ij}}\,\frac{1+3\cos\Phi_{ij}}{2}\bigg] \, ,
\quad
\Omega_{1}=\frac{\mut F^{\phi}}{R} \,\, ,
\label{eq:Omega_N}
\end{equation}
where $X_{ij}=\sqrt{2}(R/a)\sqrt{1-\cos\Phi_{ij}}$ and
$\Phi_{ij}=(2\pi/N)(j-i)$ are the respective spatial and angular
distances between particles $i$ and $j$.

\begin{figure}
\begin{indented}
\item
\includegraphics[width=0.62\textwidth]{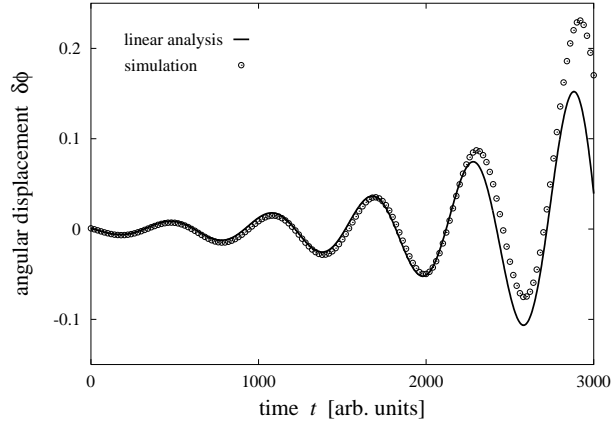}
\end{indented}
\caption{Representative example of the angular coordinate of an
unstable oscillating mode (here: $N=4$, $R/a=10$,
$\mathcal{K}=10$). The linear theory and the simulations agree 
very well up to amplitudes of several percent of $2\pi/N$. (Only every
100th data point from the simulation is plotted.)} 
\label{fig:unstable_mode}
\end{figure}

To investigate the stability of these $N$-particle clusters against
small radial and angular particle displacements, we introduce
$r_{i}=R[1+\delta\rho_{i}(t)]$ and
$\phi_{i}=\Phi_{i}(t)+\delta\phi_{i}(t)$. Using the forces
\eref{eq:vortex-force} and the Oseen tensor \eref{eq:oseen}, we
linearize the equation of motion \eref{eq:r'=mu*F} in terms of the
small perturbations $\delta\rho_{i}$ and $\delta\phi_{i}$:
\begin{eqnarray}
\fl
\frac{\d}{\d\tau}\,\delta\rho_{i}(\tau)
= &
-\mathcal{K}\,\delta\rho_{i}(\tau)
\nonumber \\
\fl
&
+\frac34\,\sum\limits_{j\neq i}\bigg[\frac{1}{X_{ij}}\bigg(
\frac{(5-3\cos\Phi_{ij})\sin\Phi_{ij}}{4(1-\cos\Phi_{ij})}
+\mathcal{K}\,\frac{1-3\cos\Phi_{ij}}{2}\bigg)
\delta\rho_{j}(\tau)\bigg]
\nonumber \\[1ex]
\fl
&
-\frac34\bigg[\sum\limits_{j\neq i}\frac{1}{X_{ij}}\,
\frac{3(1-\cos\Phi_{ij})}{4}\bigg]\delta\phi_{i}(\tau)
\nonumber \\[1ex]
\fl
&
+\frac34\sum\limits_{j\neq i}\bigg[\frac{1}{X_{ij}}\,
\frac{3(1-\cos\Phi_{ij})}{4}\,\delta\phi_{j}(\tau)\bigg]
\,\, ,
\nonumber \\[2ex]
\fl
\frac{\d}{\d\tau}\,\delta\phi_{i}(\tau)
= &
-\bigg[1+\frac34\sum\limits_{j\neq i}\frac{1}{X_{ij}}\,
\frac{3(1+3\cos\Phi_{ij})}{4}\bigg]\delta\rho_{i}(\tau)
\nonumber \\[1ex]
\fl
&
-\frac34\sum\limits_{j\neq i}\bigg[\frac{1}{X_{ij}}\bigg(
\frac{1+3\cos\Phi_{ij}}{4}+\mathcal{K}\,\frac{3\sin\Phi_{ij}}{2}\bigg)
\delta\rho_{j}(\tau)\bigg]
\nonumber \\[1ex]
\fl
&
-\frac34\sum\limits_{j\neq i}\bigg[\frac{1}{X_{ij}}\,
\frac{(7-3\cos\Phi_{ij})\sin\Phi_{ij}}{4(1-\cos\Phi_{ij})}
\,\delta\phi_{j}(\tau)\bigg]
\,\, ,
\label{eq:d/dt(delta_rho,delta_phi)}
\end{eqnarray}
where $\tau=\Omega_{1}t$ is the reduced time, and the dimensionless
parameter $\mathcal{K}=K^{r}R/F^{\phi}$ measures the strength of the
radial trap relative to the tangential driving force. The
corresponding eigenvalue problem was solved numerically by using the
computer-algebra package {\sc maple}. 
This analysis reveals that there are the following types of stable and
unstable eigenmodes (depending on the number of particles) for the
coupled displacements
$(\delta\rho_{1}(\tau),\ldots,\delta\rho_{N}(\tau),
\delta\phi_{1}(\tau),\ldots,\delta\phi_{N}(\tau))$:
(i) a constant angular shift,
(ii/iii) non-oscillating damped or unstable, and
(iv/v) oscillating damped or unstable.
\Fref{fig:unstable_mode} shows an example of an unstable oscillating
mode (type v). 
In \tref{tab:linear_modes}, we give the number of eigenmodes
classified by their eigenvalues $\lambda$ for even and odd particle
numbers. 

\begin{table}
\begin{indented}
\item \ \\[-4ex]
\item
\begin{tabular}{cccccc}
\hline\hline \\[-2.2ex]
& (i) & (ii) & (iii) & (iv) & (v) \\[1ex]
\raisebox{1.5ex}[-1.5ex]{$N$} & $\lambda=0$ &
$\lambda\in\mathbb{R}_{-}$ & $\lambda\in\mathbb{R}_{+}$ &
Re $\lambda<0$ & Re $\lambda>0$ \\[0.5ex]
\hline \\[-1ex]
odd & 1 & 1 & --- & $N-1$ & $N-1$
\\[1.5ex]
even & 1 & 2 & 1 & $N-2$ & $N-2$ \\[1.5ex]
\hline\hline
\end{tabular}
\end{indented}
\caption{Number of eigenmodes classified by their eigenvalues
$\lambda$ for even and odd particle numbers $N$. $\mathbb{R}_{\pm}$
means a positive or negative real number.
(i) constant angular shift,
(ii/iii) non-oscillating damped or unstable modes, (iv/v) oscillating
damped or unstable modes.}
\label{tab:linear_modes}
\end{table}

The occurence of stable and unstable modes can be compared to a
saddle point in the framework of the analysis of dynamic systems. A
cluster configuration with arbitrary small radial and angular
displacements of the particles is practically unstable since
contributions of unstable modes will grow, while stable modes will
relax to zero.

For $N=3$, we have also determined the eigenvalues analytically by an
expansion to first order in $a/R$: 
\begin{eqnarray*}
\lambda_{1} &= 0
& \mbox{(type i)} \, ,
\\[1ex]
\lambda_{2} &= -\mathcal{K}\bigg(1-\frac{\sqrt{3}}{10}\,
\frac{a}{R}\bigg)
& \mbox{(type ii)} \, ,
\\[1ex]
\lambda_{3,4} &= -\mathcal{K}-\bigg[\bigg(\frac{\sqrt{3}}{20}
+\frac{27\sqrt{3}}{32}\,\frac{1}{\mathcal{K}}\bigg)
\pm\i\,\frac{13\sqrt{3}}{32}\,\bigg]\frac{a}{R}
\qquad & \mbox{(type iv)} \, ,
\\[1ex]
\lambda_{5,6} &= \bigg(\frac{27\sqrt{3}}{2}\,\frac{1}{\mathcal{K}}
\pm\i\,\frac{17\sqrt{3}}{32}\,\bigg)\frac{a}{R}
& \mbox{(type v)} \, .
\end{eqnarray*}
In this case, a non-oscillating unstable mode (type iii) does not
exist; it only occurs for even $N$ (see \tref{tab:linear_modes}).

%%% DYNAMICS %%%%%%%%%%%%%%%%%%%%%%%%%%%%%%%%%%%%%%%%%%%%%%%%%%%%%%%%%%%%%%%%%%

\section{Nonlinear dynamics of perturbed clusters}
\label{sec:nonlinear_dynamics}

\begin{figure}
\begin{indented}
\item
\includegraphics[width=0.81\textwidth]{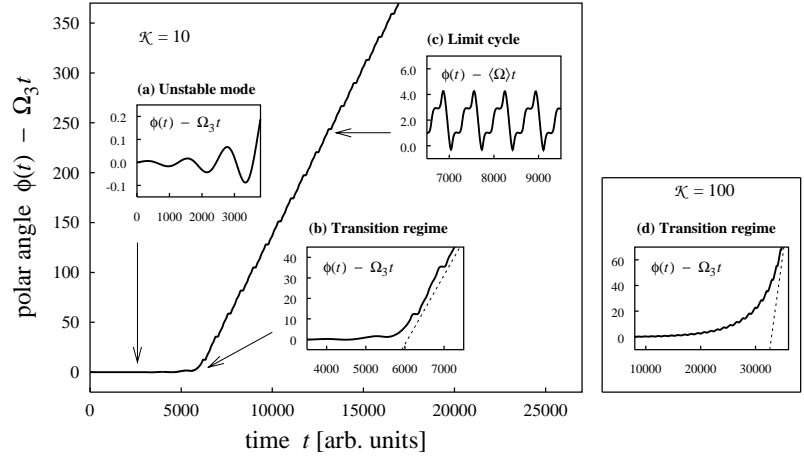}
\end{indented}
\caption{Dynamic transition in the angular coordinate from an unstable
linear mode to the periodic limit cycle for three particles
($R/a=10$). The main curve and the insets (a), (b), and (d) show the
angle $\phi(t)$ relative to $\Omega_{N}t$ as a function of time.
In the inset (c), $\phi(t)-\langle\Omega\rangle t$ is plotted.
Plots (a)--(c) correspond to a stiffness of $\mathcal{K}=10$.
For comparison with (b), the transition regime for $\mathcal{K}=100$ is
shown in (d). The dashed lines in (b) and (d) indicate the limit
cycles.}
\label{fig:dynamic_transition}
\end{figure}

The amplitude of an unstable mode grows up to a certain
magnitude in agreement with the linear analysis, until the
nonlinear dynamics takes over (see \fref{fig:unstable_mode}).
The amplitudes saturate, and the system finally tends towards a
periodic limit cycle with oscillating particle distances.
In \fref{fig:dynamic_transition} we show an example of such a dynamic
transition from an unstable linear mode to the periodic limit cycle by
plotting the angular coordinate $\phi(t)$ relative to $\Omega_{N}t$ as
a function of time. Note that $\Omega_{N}$ is the rotational frequency
of the regular $N$-particle cluster introduced in \eref{eq:Omega_N}.

The mean slope in each of the two regimes (linear mode and limit
cycle) gives a well-defined mean orbital frequency
$\langle\Omega\rangle$, which is $\Omega_{N}$ for the linear
mode. Clearly, $\langle\Omega\rangle$ and therefore the mean velocity
of the particles is increased by the transition to the limit cycle,
which means that the mean drag force on the particles is reduced.

\begin{figure}
\begin{indented}
\item
\includegraphics[width=0.81\textwidth]{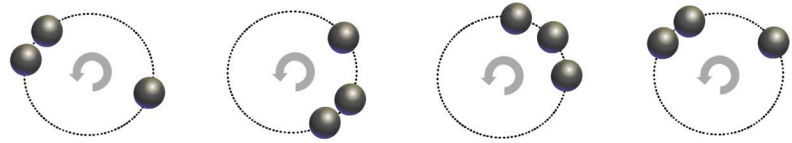}
\end{indented}
\caption{Basic mechanism of the periodic limit cycle. Two particles in
close contact move faster than a single particle. When this pair
reaches the third particle, they form an intermediate triplett, and
finally, the first two particles escape as a new pair.}
\label{fig:limit-cycle-mechanism}
\end{figure}

The character of the transition depends on the trap stiffness. It is
quite sharp for weaker traps, where the transition takes place within
a few oscillations, as shown in \fref{fig:dynamic_transition}(b).
For stronger traps, the mean frequency increases smoothly from the
linear regime to the limit cycle, which is illustrated in
\fref{fig:dynamic_transition}(d). 
Furthermore, the onset of the transition is shifted to later times
when the trap becomes stiffer.

\Fref{fig:limit-cycle-mechanism} introduces the basic mechanism
underlying the periodic limit cycle. Two particles in close contact
move faster than a single particle since the friction per particle is
reduced due to the well-known effect of drafting.
When such a pair reaches the third particle, they form a triplett for
a short time. In this configuration, the mobility of the middle
particle is larger since it is ``screened'' from the fluid flow by the
outer particles. It therefore pushes the particle in front, so that
the first two particles ``escape'' from the last one.
The same principle also holds for more than three particles. In these
cases, there can be more than one pair of particle.

\begin{figure}
\begin{indented}
\item
\includegraphics[width=0.81\textwidth]{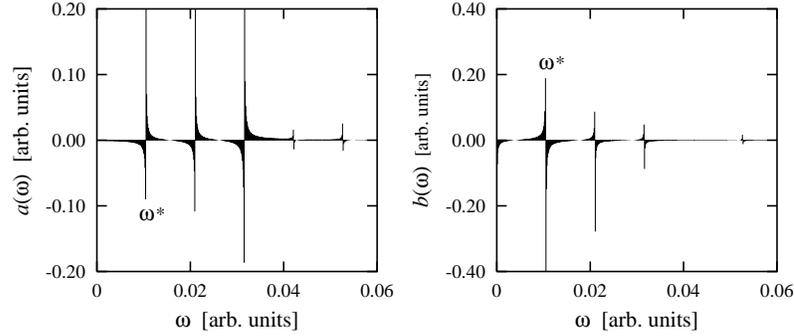}
\end{indented}
\caption{Fourier spectrum [$\cos$-coefficients $a(\omega)$ and
$\sin$-coefficients $b(\omega)$] of the periodic limit cycle of three
particles. Note that the frequencies $\omega$ are defined
relative to the mean orbital frequency $\langle\Omega\rangle$.}
\label{fig:fourier-spectrum}
\end{figure}

For the limit cycle of three particles, we have performed a harmonic
analysis using Fast Fourier Transformation. In order to
separate the fast orbital dynamics of the particles 
(characterized by the mean angular frequency $\langle\Omega\rangle$)
from their relative motions, we calculate the Fourier transform of
$\phi(t)-\langle\Omega\rangle t=\sum_{\omega}[a(\omega)\cos\omega
t+b(w)\sin\omega t]$. This yields the fingerprint of the dynamics 
relative to the mean circling velocity. In
\fref{fig:fourier-spectrum}, we show the corresponding Fourier
spectrum. Besides the characteristic frequency $\omega^{*}$ of the
limit cycle, the higher harmonics $2\omega^{*}$ and $3\omega^{*}$ are
also very pronounced.

\begin{figure}
\begin{indented}
\item
\includegraphics[width=0.48\textwidth]{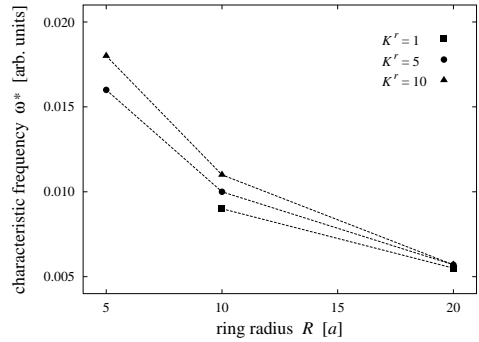}
\end{indented}
\caption{Characteristic frequency $\omega^{*}$ of the periodic limit
cycle of three particles versus ring radius $R$. The different curves
are related to different trap stiffnesses $K^{r}$.}
\label{fig:characteristic_frequency}
\end{figure}

The characteristic frequency $\omega^{*}$ decreases with increasing
ring radius, as shown in \fref{fig:characteristic_frequency}, because
it takes longer for a particle pair to reach the third particle on a
larger ring. 
At constant ring radius, the characteristic frequency increases with
the trap stiffness since particles in a stronger trap are better aligned 
along the ring, which makes the mechanism of the limit cycle
more effective.

If the strength of the radial trap is decreased, the radial 
displacements of the particles increase so that they can pass each other. 
This happens at a reduced trap stiffness $\mathcal{K}=K^{r}R/F^{\phi}$ of the
order of 1. For three particles and four particles, the limit cycle
then consists of a compact triangular or rhombic-shaped cluster
circling and rotating in the trap.

\begin{figure}
\includegraphics[width=\textwidth]{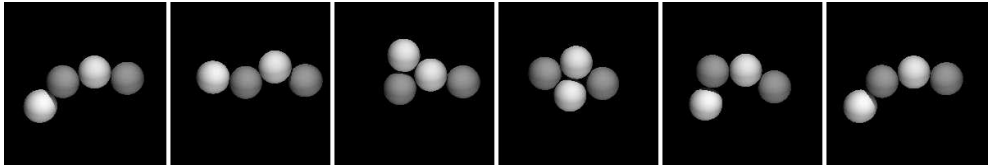}
\caption{Dynamics of four particles in a weak radial trap. The motion
shown is relative to the mean rotation along the ring trap. Only the
relevant part of the ring is shown, the center of the ring is located
at the bottom of the snapshots. The particles are running from right
to left. During one cycle, the first and the third particle are
exchanged. The particles are shaded differently in order to
distinguish them.}
\label{fig:worm-dynamics}
\end{figure}

For four particles, we also find the limit cycle illustrated in
\fref{fig:worm-dynamics}. It occurs if the radial trap has medium 
strength so that compact particle clusters only have a finite life time.
The first particle on the left is pushed in outward radial direction 
by the particles behind it and subsequently passed by the second particle. 
It, then, pushes the third one in inward radial direction, which
in turn takes over the lead of the chain. Finally, the initial state is
rebuild, where the first and the third particle are exchanged. Note
that there is an intermediate state where the particles form a 
compact rhombic-shaped cluster, which, however, is not stable.

%%% VELOCITIES %%%%%%%%%%%%%%%%%%%%%%%%%%%%%%%%%%%%%%%%%%%%%%%%%%%%%%%%%%%%%%%%

\section{Particle velocities}
\label{sec:velocities}

\begin{figure}
\includegraphics[width=0.49\textwidth]{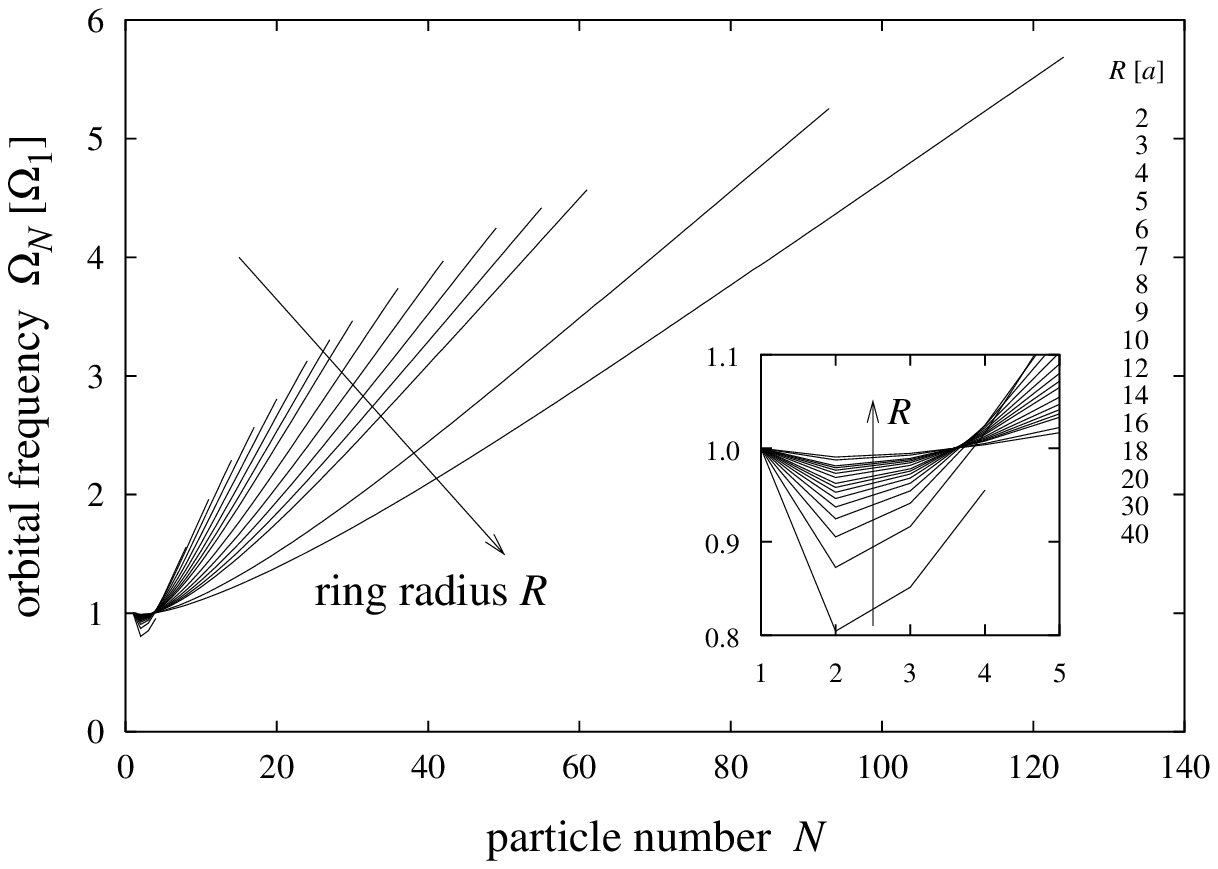}
\hfill
\includegraphics[width=0.49\textwidth]{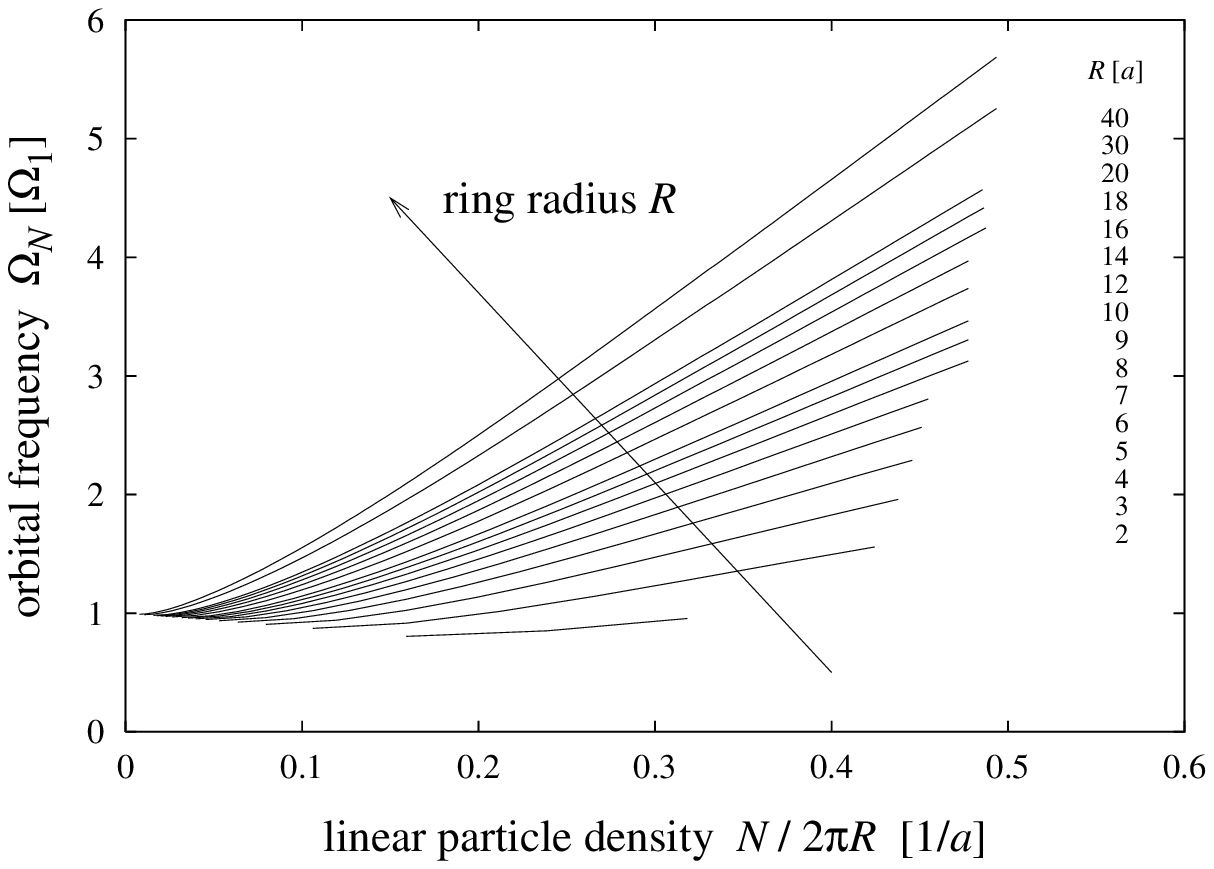}
\caption{Circling frequencies of regular $N$-particle clusters,
normalized to the frequency of a single sphere, as obtained from 
simulations. The two plots show the same data, left: as a function of
particle number, right: as a function of particle line density.
The arrow gives the direction of increasing ring radius.}
\label{fig:cluster-velocities}
\end{figure}

In \fref{fig:cluster-velocities} (left), we study the circling
frequencies of regular clusters as a function of particle number $N$
for different ring radii $R$. At constant radius, the frequency
increases with the number of particles since then the particles are
closer together which reduces the drag. Note that for less than four
particles, the frequency is reduced relative to the single particle
value. Here, hydrodynamic interactions across the ring increase the drag
on the particles. This effect is clearly more pronounced for small rings.

At constant particle distance $2\pi R/N$ or constant line density 
of the particles $N/(2\pi R)$, the drag is reduced when the ring
radius increases [see \fref{fig:cluster-velocities} (right)]. This is
due to the fact that in a ring with smaller curvature the particles
are better ``aligned'' behind each other.

\begin{figure}
\begin{indented}
\item
\includegraphics[width=0.49\textwidth]{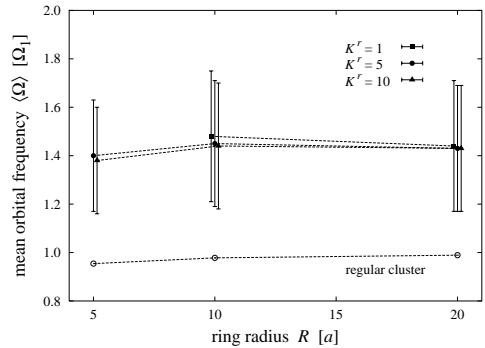}
\end{indented}
\caption{Mean circling frequencies of three particles in the periodic
limit cycle, normalized to the frequency of a single sphere. The upper
three curves are plotted for different trap stiffnesses $K^{r}$. For
comparison, the lower curve shows the values for the regular 
three-particle cluster. The errorbars indicate the range within which 
the orbital velocities vary in the limit cycle.}
\label{fig:limit-cycle-velocity}
\end{figure}

In the limit cycle, discussed in the previous section, the mean 
orbital frequency is larger compared to the velocity of the 
corresponding regular configuration. The quantitative effect
is illustrated in \fref{fig:limit-cycle-velocity}. We attribute this
to the drafting of particle doublets which obviously reduces their 
drag force. We observe that the dependence of the limit-cycle velocity on
the trap stiffness is weak, in contrast to the characteristic frequency
of the angular displacements, as already discussed in 
\fref{fig:characteristic_frequency}.

%%% CONCLUSION %%%%%%%%%%%%%%%%%%%%%%%%%%%%%%%%%%%%%%%%%%%%%%%%%%%%%%%%%%%%%%%%

\section{Conclusion}
\label{sec:conclusion}

We modelled the circling of particles in an optical
vortex and illustrated how hydrodynamic interactions govern the
nonlinear dynamics of the coupled particle motion. We hope that
our theoretical investigation initiates a detailed study within
experiments. Possible extensions of our work concern the tangential
driving force which could be modulated along the ring or which could
possess stochastic contributions.

%%% ACKNOWLEDGMENTS %%%%%%%%%%%%%%%%%%%%%%%%%%%%%%%%%%%%%%%%%%%%%%%%%%%%%%%%%%%

\ack

We are grateful to Erwin Frey who initiated this work by pointing out
the experiments of Jennifer Curtis and David Grier. Furthermore, we
would like to thank Jennifer Curtis for stimulating discussions.
This work was supported by the Deutsche For\-schungs\-ge\-mein\-schaft
through the Sonderforschungsbereich Transregio 6 ``Physics of colloidal
dispersions in external fields''. H. S. acknowledges financial support
from the Deutsche Forschungsgemeinschaft by Grant No. Sta 352/5-1.

%%% REFERENCES %%%%%%%%%%%%%%%%%%%%%%%%%%%%%%%%%%%%%%%%%%%%%%%%%%%%%%%%%%%%%%%%

\end{document}